\documentclass[twocolumn]{aastex631} 

\usepackage{amsmath}
\usepackage{soul}
\usepackage{booktabs}
\usepackage{multirow}

\usepackage{xspace}
\newcommand{\PS}{Press-Schechter\xspace}
\newcommand{\ST}{Sheth-Tormen\xspace}
\newcommand{\hmpc}{$h^{-1}\mathrm{Mpc}$\xspace}
\newcommand{\hmsun}{$h^{-1}M_\odot$\xspace}
\newcommand{\lcdm}{$\Lambda$CDM\xspace}
\newcommand{\vide}{\tt VIDE\normalfont\xspace}
\newcommand{\zobov}{\tt ZOBOV\normalfont\xspace}

\begin{document}

\title{The halo bias inside cosmic voids}

\correspondingauthor{Giovanni Verza}
\email{giovanni.verza@pd.infn.it}

\author[0000-0002-1886-8348]{Giovanni Verza}
\affiliation{INFN, Sezione di Padova,
via Marzolo 8, I-35131 Padova, Italy}
\affiliation{Dipartimento di Fisica e Astronomia ``G. Galilei",
Universit\`a degli Studi di Padova, via Marzolo 8, I-35131 Padova, Italy}

\author[0000-0003-0125-3563]{Carmelita Carbone}
\affiliation{INAF – Istituto di Astrofisica Spaziale e Fisica cosmica di Milano (IASF-MI), Via Alfonso Corti 12, I-20133 Milano, Italy}

\author[0000-0001-9856-1970]{Alessandro Renzi}
\affiliation{Dipartimento di Fisica e Astronomia ``G. Galilei",
Universit\`a degli Studi di Padova, via Marzolo 8, I-35131 Padova, Italy}
\affiliation{INFN, Sezione di Padova,
via Marzolo 8, I-35131 Padova, Italy}

\begin{abstract}
\noindent
The bias of dark matter halos and galaxies is a crucial quantity in many cosmological analyses. In this work, using large cosmological simulations, we explore the halo mass function and halo bias within cosmic voids. For the first time to date, we show that they are scale-dependent along the void profile, and provide a predictive theoretical model of both the halo mass function and halo bias inside voids, recovering for the latter a 1\% accuracy against simulated data. These findings may help shed light on the dynamics of halo formation within voids and improve the analysis of several void statistics from ongoing and upcoming galaxy surveys.
\end{abstract}

\section{Introduction}

Recently, many cosmological statistics of cosmic voids have been explored, as for instance the void size function and void clustering which can probe the underlying cosmological model of the Universe~\citep{pisani_2015_abundance,hamaus_2015,sahlen_2016,cai_2016,chuang_2017,achitouv_2017,hawken_2017,hamaus_2017,sahlen_2018,achitouv_2019,nadathur_2019_BOSS,kreisch_2019,verza_2019,pisani_2019,hamaus_2020,hawken_2020,nadathur_2020,aubert_2020,bayer_2021,kreisch_2021,moresco_2022_cosmological_probes,woodfinden_nadathur_2022,euclid_vsf}. 

Past, ongoing and upcoming spectroscopic and photometric galaxy surveys, such as 6dFGS~\citep{jones_2009_6DF}, VIPERS~\citep{guzzo_2014_VIPERS}, BOSS~\citep{dawson_2013_BOSS}, eBOSS~\citep{dawson_2016_eBOSS}, DES~\citep{DES_2016}, DESI~\citep{DESI_2016}, PSF~\citep{PFS_2016}, Euclid~\citep{laureijs_20211_euclid_report}, the Roman Space Telescope~\citep{spergel_2015_WFIRST}, SPHEREx~\citep{dore_2018_SPHEREx}, and the Vera Rubin Observatory~\citep{ivezic_2019_LSST}, are elevating cosmic voids to an effective and competitive new cosmological probe helping in testing cosmological scenarios and theories of structure formation.

Cosmic voids are large underdense regions in the large-scale structure of the Universe, span a large range of scales and constitute the largest observable objects in the Universe. Their size and underdense nature make them particularly suited to probe dark energy and modified gravity~\citep{lee_2009,lavaux_wandelt_2009,biswas_2010,li_2010_MG,clampitt_2013_MG,spolyar_2013,cai_2015,pisani_2015_abundance,zivick_2015,pollina_2015,achitouv_2016,sahlen_2016,falck_2018,sahlen_2018,paillas_2019,perico_2019,verza_2019,contarini_2021,euclid_vsf}, massive neutrinos~\citep{massara_2015,banerjee_2016,kreisch_2019,sahlen_2019,schuster_2019,zhang_2020, kreisch_2021,euclid_vsf}, primordial non-Gaussianity~\citep{chan_2019}, and physics beyond the standard model~\citep{peebles_2001,reed_2015,yang_2015,baldi_2016,lester_2021,arcari_2022}.

Crucial quantities for many cosmological analyses involving cosmic voids are the halo and galaxy distributions within their volume and their relation with the underlying matter density field \citep{sutter_2014_DMgVoids, Quilis_2021}.
Therefore, in this work we use large cosmological simulations to investigate the halo mass function (HMF) and bias inside voids. We show that, for such halos, the mass function is not universal along the void profile, rather it depends on the distance from the void center, and we provide a theoretical model describing it. Then, we show that the knowledge of the behavior of the HMF as a function of such distance allows us to accurately model the halo bias inside voids, going beyond the linear parameterization widely used in the literature. This opens a window on the physics of halo formation within cosmic voids.

\section{Simulations and void finder}
For this study we use the “Dark Energy and Massive Neutrino Universe” (\href{https://www.researchgate.net/project/DEMN-Universe-DEMNUni}{DEMNUni}) set of simulations~\citep{carbone_2016_demnuni}.
These simulations are characterized by a comoving volume of $(2 \: h^{-1}\mathrm{Gpc})^3$, big enough to include the very large-scale perturbation modes, filled with $2048^3$ dark matter particles and, when present, $2048^3$ neutrino particles. 
The simulations have a Planck 2013~\citep{planck2013} baseline \lcdm reference cosmology and flat geometry. The DEMNUni set is composed by 15 simulations, implementing the cosmological constant and 4 dynamical dark energy equations of state for each of the total neutrino mass considered, i.e. $\sum m_\nu = 0,\, 0.16,\, 0.32\, {\rm eV}$.

Dark matter halos are identified using a friends-of-friends (FoF) algorithm~\citep{davis_1985_fof} applied to dark matter particles, with a minimum number of particles fixed to 32, corresponding to a mass of $\sim 2.5 \times 10^{12} h^{-1}M_\odot$, and a linking length of 0.2 times the mean particle separation. FoF halos are further processed with the {\sc subfind} algorithm~\citep{springel_2001_gadeget,dolang_2009_gadget} to get subhalo catalogs. With this procedure, some of the initial FoF parent halos are split into multiple subhalos, with the result of an increase of the total number of identified objects and of a lower minimum mass limit. The minimum number of particles for a subhalo is set to 20, corresponding to a minimum mass of $\sim 1.7 \times 10^{12} h^{-1}M_\odot$. Note that the {\sc subfind} algorithm is a halo estimator, therefore in the following with the term ``halo'' we will refer to the objects identified by {\sc subfind} algorithm.

To identify voids and build the void catalogs we use the second version of the ``Void IDentification and Examination'' (\href{http:www.cosmicvoids.net}{\vide}) public toolkit~\citep{sutter_2015_vide}.
\vide is based on the tessellation plus watershed void finding technique~\citep{platen_2007} with the Delaunay tessellation/DTFE~\citep{schaap_2000_dtfe} implemented in \zobov~\citep{neyrinck_2008} to estimate the density field. 
The algorithm groups nearby Voronoi cells into zones, corresponding to local catchment ``basins", that are identified as voids; the void boundary is the watershed of the basin, i.e. the ridge of relative overdensity that surrounds it.

For the present analysis, we consider particle snapshots from the DEMNUni set in the \lcdm cosmology at redshifts $z=0$, 0.49, 1.05. For each of them, we build a catalog of halo-traced voids detected in the halo distributions with three different minimum masses: $2.5 \times 10^{12}$\hmsun, $10^{13}$\hmsun, $2.5 \times 10^{13}$\hmsun.
In this work we consider only voids traced by halos, therefore, hereafter with the term ``voids'' we will refer only to halo-traced voids.

We characterize the detected voids according to two parameters reconstructed by \vide: 
\begin{itemize}
\item The void size, measured via the effective radius, i.e. the radius of the sphere with the same volume of the considered void, $R_{\rm eff} =(3/4 \pi) \left[  \sum_i V_i \right]^{1/3}$, with $V_i$ the volume of the $i^{\rm th}$ Voronoi cell building up the void;
\item The void center, estimated as the barycenter by weighting the volumes of the contributing Voronoi cells. 
\end{itemize}
Note that \vide detects all the relative minima in the tracer (halo) distribution. 
Therefore, in order to restrict our analysis to true underdensities, for each tracer (i.e. halos with different minimum masses) we consider only voids for which the tracer density, averaged over a sphere of radius $R_\mathrm{eff}/4$, is less than the mean tracer density in the simulated comoving snapshot (box) at the redshift considered. Moreover, to avoid contamination from resolution effects~\citep{verza_2019,euclid_vsf}, we show our analysis only for voids with radius, $R_{\rm eff}$, larger than 2.5 the mean halo separation (mhs), i.e. $R_{\rm eff} \geq 2.5 \times {\rm mhs}$, where ${\rm mhs}\equiv(N_{\rm h}/V)^{-1/3}$ and $N_{\rm h}$ is the total number of halos in a simulated box of volume $V$. In Tab.~\ref{tab:halo_mps} we list the different mhs for each of the simulated boxes and halo catalogs considered, together with the corresponding minimum halo masses.
\begin{table}[t!]
\centering
\begin{tabular}{cccc}
\toprule
\multirow{2}{*}{$\quad z \quad$} & \multicolumn{3}{c}{$M_{\rm min} \quad [h^{-1}M_\odot]$} \\
&  $ \quad 2.5 \times 10^{12} \quad $ & $ \quad 10^{13}  \quad $ & $ \quad 2.5 \times 10^{13}  \quad$\\
\tableline
0  &  7.9  &  12.3  &  17.3 \\ 
0.49  &  8.2  &  13.4  &  19.8 \\ 
1.05   &  8.9  &  15.7  &  25.5 \\ 
\midrule
\bottomrule
\end{tabular}
\caption{The second, third and fourth columns show the mean halo separation in \hmpc unit measured at each redshift, $z$, and for each minimum halo mass, $M_{\rm min}$, considered.}
\label{tab:halo_mps}
\end{table}

\section{Void density profile}
The void density profile is defined as the mean density contrast in spherical shells at a distance $r$ from the void center~\citep{hamaus_2014_profiles}. Note that this definition, usually known as stacked void density profile~\citep{lavaux_2012_AP}, is an estimation of the void-tracer cross-correlation~\citep{hamaus_2017,hamaus_2020}. In the averaging procedure, we normalize the distance from the void center with respect to $R_{\rm eff}$. In this way, the characteristic shape of the void profile is retained in the result of the procedure~\citep{hamaus_2017}. 
We measure the void density profile in the halo distribution and in the dark matter distribution within the corresponding halo-traced voids. In order to measure the void profiles, we consider spherical shells, $\mathcal{S}^{\rm v}(r)$, of thickness $\Delta r=0.067 R_{\rm eff}$, spanning a radial distance from the void center in the range $0<r<4R_{\rm eff}$. Finally, to estimate the uncertainty on the resulting stacked density profiles we use the jackknife method~\citep{miller_1974_jackknife}.

\begin{figure}[t!]
\centering
\includegraphics[width=\columnwidth]{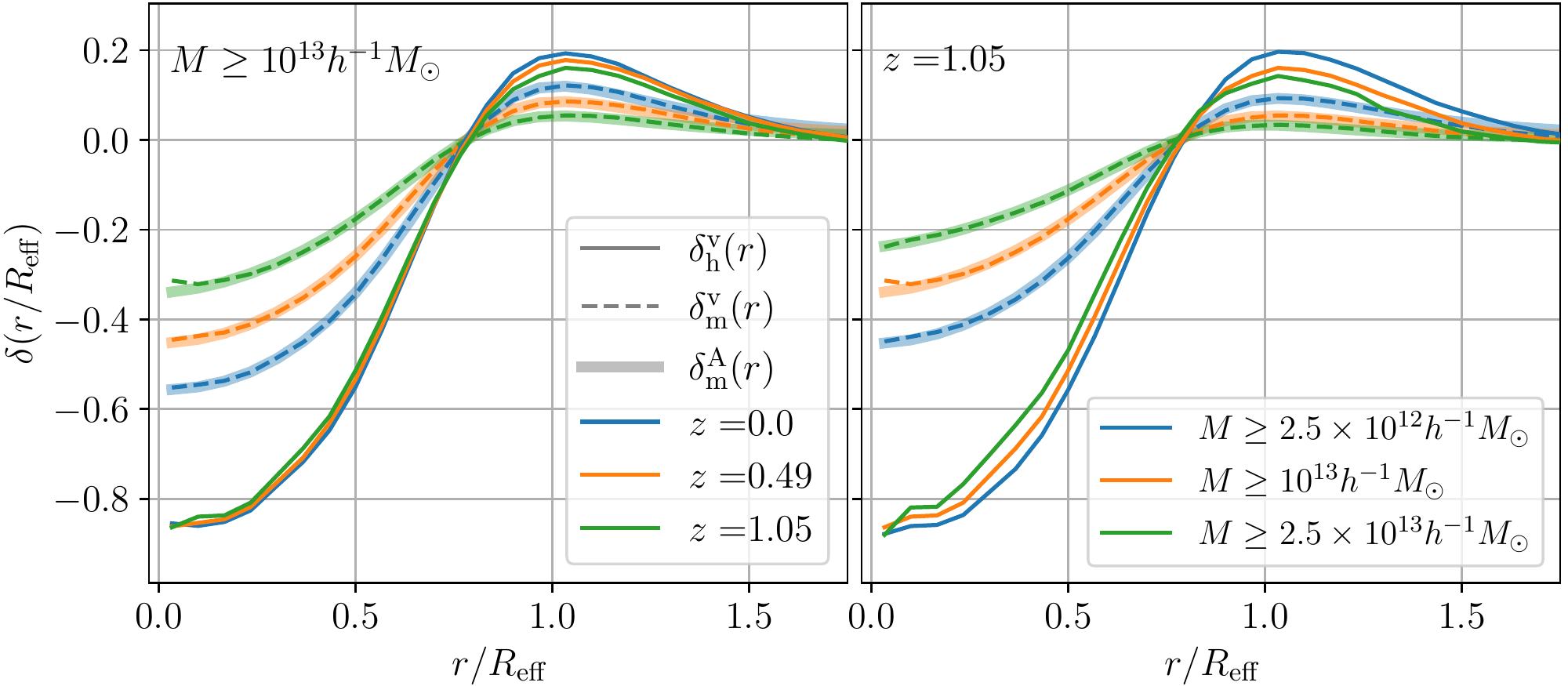}
\caption{Stacked density profiles for voids traced by halos with $M \geq 10^{13} h^{-1}M_\odot$ at redshifts $z=0$, 0.49, 1.05 (left), and for various minimum halo masses, $M=2.5 \times 10^{12}$\hmsun, $10^{13}$\hmsun, $2.5 \times 10^{13}$\hmsun, at $z=1.05$ (right). The solid lines show the density profiles in the halo distributions, $\delta_{\rm h}^{\rm v}(r)$, the dashed lines in the underlying matter distribution, $\delta_{\rm m}^{\rm v}(r)$, within the halo-traced voids. The transparent lines show the analytical profile, $\delta_{\rm m}^{\rm A}(r)$,  of \citet{hamaus_2014_profiles} fitted against $\delta_{\rm m}^{\rm v}(r)$.}
\label{fig:profile}
\end{figure}

Fig.~\ref{fig:profile} shows the void density profiles measured in the halo distributions, $\delta_{\rm h}^{\rm v}(r)$, and in the underlying matter distributions, $\delta_{\rm m}^{\rm v}(r)$. In particular, the left panel shows the profiles of voids traced by halos with $M\geq 10^{13}h^{-1}M_\odot$ at various redshifts, and the right panel shows the voids profiles at redshift $z=1.05$ for different minimum halo masses.
Both $\delta_{\rm h}^{\rm v}(r)$ (solid lines) and $\delta_{\rm m}^{\rm v}(r)$ (dashed lines) show the same qualitative features: the void density profiles are negative at $r/R_{\rm eff}\sim 3/4$ where they present a sign inversion, then peak at about $r/R_{\rm eff} \sim 1$, and finally, at larger distances, approach the mean density of the considered tracers (halos or matter).
The peak is called ``compensation wall'' and corresponds to the overdensities of the watershed, therefore is located at about $r/R_{\rm eff} \sim 1$~\citep{hamaus_2014_profiles}. Note that, for each minimum halo mass considered, the two void profiles have different (scale dependent) shapes, due to the effective halo bias which increases with the halo mass. Finally, the transparent lines show the analytical matter density profiles $\delta^{\rm A}_{\rm m}(r)$ of Eq.~(2) in~\citet{hamaus_2014_profiles}, where the parameters $\delta_c$, $r_v$, $\alpha$ and $\beta$ have been fitted against measurements of $\delta_{\rm m}^{\rm v}(r)$ in the simulations\footnote{The quantity $\delta_c$ in~\citet{hamaus_2014_profiles} refers to the void central density and it should not be confused with the linear density contrast of spherical collapse, discussed in Sec.~\ref{sec:HFM_theo}. Analogously for the $\alpha$ and $\beta$ parameters in~\citet{hamaus_2014_profiles}.}.

\section{Halo mass function in voids}\label{sec:HFM_theo}
From a theoretical point of view, the availability of a model for the HMF within voids automatically provides a way to predict the halo bias inside voids.
Indeed, the excursion-set formalism~\citep{Peacock1990, bond_et_al1991, Lacey_Cole1993} provides a framework to obtain a HMF which depends only on two quantities: i) the variance of linear matter perturbations as a function of the halo mass scale, $M$, and ii) the linear density contrast threshold, i.e. the critical overdensity required for matter structure virialization at redshift $z$, which, in case of non-spherical collapse, depends on the mass $M$ and is called the ``moving barrier"~\citep{press_schechter_1974,epstein_1983,Peacock1990,bond_et_al1991,mo_white_1996,sheth_tormen_2001,sheth_tormen_2002}:
\begin{align}\label{barrier}
B(\sigma^2,z)=\sqrt{q}\delta_{\rm sc}(z)\left[1+\beta(\alpha\nu)^{-\alpha}\right] \;.
\end{align}
Here $\nu\equiv\delta_{\rm sc}(z)/\sigma(M)$, $\sigma(M)$ is the rms value of the initial density fluctuation field, filtered on a scale $R(M)$, and evolved using linear theory to the present time, while $\delta_{\rm sc}(z)$ is the critical overdensity required for \emph{spherical} collapse at $z$, extrapolated to the present time via the linear growth factor, $D$, of density fluctuations: $\delta_{\rm sc}(z)=\delta_{\rm c}\,D(z=0)/D(z)$, with $\delta_{\rm c}\approx3\,(12\pi)^{2/3}(1+0.0123\log_{10}\Omega_{\rm m})/20$, $\Omega_{\rm m}=\Omega_{0\rm m}(1+z)^3/E^2(z)$, and $E(z)=H(z)/H_0$, being $H(z)$ the Hubble parameter, $\Omega_{\rm m}$ the matter density parameter, and the $0$-subscript representing their today values. The $\alpha$ and $\beta$ parameters come from the ellipsoidal dynamics, and the $q$ value from the normalization of the model to cosmological simulations \citep{sheth_tormen_2002}.

Using Eq.~\eqref{barrier} in the excursion-set approach, in order to obtain the distribution of the first crossings of the barrier by independent random walks, \citet{sheth_tormen_1999}, \citet{sheth_tormen_2001}, and \citet{sheth_tormen_2002} derived the average comoving number density of halos with mass in the range $(M, M + {\rm d} M)$, i.e. the so-called unconditional HMF:
\begin{equation}\label{eq:universal_HMF}
\frac{{\rm d} n_{\rm h}}{{\rm d} M}= \frac{\bar{\rho}_{\rm m}}{M} f[\nu(z),p,q] \left| \frac{{\rm d} \nu(z)}{{\rm d} M}\right|\,,
\end{equation}
where $\bar{\rho}_{\rm m}$ is the mean comoving mass density of the Universe. Here the multiplicity function, $f$, is the fraction of fluctuations of mass $M$ that collapsed in a halo, and reads
\begin{equation}\label{eq:ST_multiplicity}
f(\nu,p,q) =  \sqrt{\frac{2}{\pi}} A \left[ 1 + (q \nu^2)^{-p} \right] \sqrt{q} e^{ -q\nu^2/2},
\end{equation}
where $A=\left[ 1 + \Gamma \left( 1/2 -p \right)/(2^p\sqrt{\pi}) \right]^{-1}$ is the normalization factor, and $p$ and $q$ parameterize the so-called ``homogeneous ellipsoid collapse model''. However, their dependence on the collapse dynamics is extremely complex to be derived {\it a priori}, therefore they are usually fitted to the HMF measured in large cosmological simulations~\citep{sheth_tormen_2002}. 
Note that for $p=0$ and $q=1$ one recovers the \PS multiplicity function for the spherical halo collapse~\citep{press_schechter_1974,sheth_tormen_2002}.
\begin{figure*}[t!]
\centering
\includegraphics[width=\textwidth]{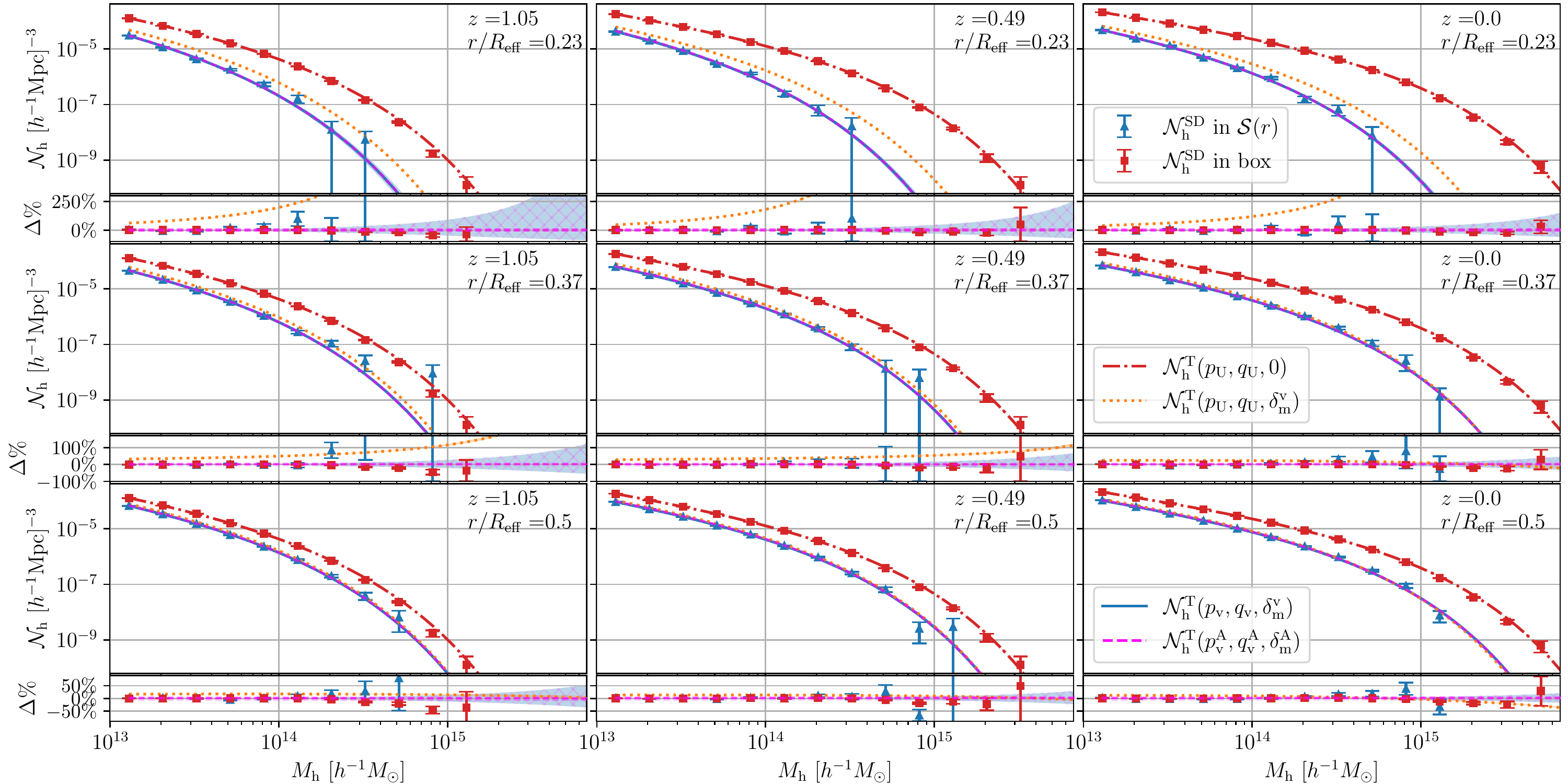}
\caption{Differential number density ${\cal N}_{\rm h}$ of halos at $z=1.05,\,0.49,\,0$ (from left to right). Symbols represent measurements in the entire comoving box (red squares) and in spherical shells ${\cal S}(r)$, (blue triangles), at $r/R_{\rm eff} = 0.23,\,37,\,0.5$ (up to down) for voids detected in the halo distribution with $M \geq 10^{13} h^{-1}M_\odot$. Errorbars show the Poissonian uncertainty. Lines show the theoretical models ${\cal N}_{\rm h}^{\rm T}$ as discussed in the text: ${\cal N}_{\rm h}^{\rm T}(p_{\rm U},q_{\rm U},0)$ is case i), i.e. ${\cal N}_{\rm h}^{\rm T}$ via Eq.~\eqref{eq:universal_HMF} (red dash-dotted lines); ${\cal N}_{\rm h}^{\rm T}(p_{\rm U},q_{\rm U},\delta^{\rm v}_{\rm m})$ is case ii), i.e. ${\cal N}_{\rm h}^{\rm T}$ via Eq.~\eqref{eq:eulerian_HMF} (orange dotted lines); ${\cal N}_{\rm h}^{\rm T}(p_{\rm v},q_{\rm v},\delta^{\rm v}_{\rm m})$ is case iii), i.e. ${\cal N}_{\rm h}^{\rm T}$ via Eq.~\eqref{eq:HMF_voids} (blue solid lines); ${\cal N}_{\rm h}^{\rm T}(p^{\rm A}_{\rm v},q^{\rm A}_{\rm v},\delta^{\rm A}_{\rm m})$ is case iv), i.e. ${\cal N}_{\rm h}^{\rm T}$ via Eq.~\eqref{eq:HMF_voids} with $p^{\rm A}_{\rm v}(r)$, $q^{\rm A}_{\rm v}(r)$, and $\delta^{\rm A}_{\rm m}(r)$ (magenta dashed lines). Shaded areas show the uncertainty on the $p$ and $q$ parameters. The subplots in each panel show the percent differences of: ${\cal N}_{\rm h}^{\rm SD}$ measured in the comoving box with respect to ${\cal N}_{\rm h}^{\rm T}(p_{\rm U},q_{\rm U},0)$, i.e. case i) (red squares); ${\cal N}_{\rm h}^{\rm SD}$ measured in ${\cal S}(r)$ with respect to ${\cal N}_{\rm h}^{\rm T}(p_{\rm v},q_{\rm v},\delta^{\rm v}_{\rm m})$, i.e. case iii) (blue triangles); ${\cal N}_{\rm h}^{\rm T}(p_{\rm U},q_{\rm U},\delta^{\rm v}_{\rm m})$, i.e. case ii), with respect to ${\cal N}_{\rm h}^{\rm T}(p_{\rm v},q_{\rm v},\delta^{\rm v}_{\rm m})$, i.e. case iii) (orange dotted lines); ${\cal N}_{\rm h}^{\rm T}(p^{\rm A}_{\rm v},q^{\rm A}_{\rm v},\delta^{\rm A}_{\rm m})$, i.e. case iv), with respect to ${\cal N}_{\rm h}^{\rm T}(p_{\rm v},q_{\rm v},\delta^{\rm v}_{\rm m})$, i.e. case iii) (magenta dashed lines). Blue shaded areas show the relative uncertainties of ${\cal N}_{\rm h}^{\rm T}(p_{\rm v},q_{\rm v},\delta^{\rm v}_{\rm m})$, i.e. case iii) due to $q_{\rm v}$ and $p_{\rm v}$ errors; magenta hatched areas show the relative uncertainties of ${\cal N}_{\rm h}^{\rm T}(p^{\rm A}_{\rm v},q^{\rm A}_{\rm v},\delta^{\rm A}_{\rm m})$, i.e. case iv), with respect to ${\cal N}_{\rm h}^{\rm T}(p_{\rm v},q_{\rm v},\delta^{\rm v}_{\rm m})$, i.e. case iii) due to $q^{\rm A}_{\rm v}$ and $p^{\rm A}_{\rm v}$ errors.}
\label{fig:HMF_in_shell}
\end{figure*}
Once one has a model for the HMF, the halo distribution and halo bias can be modeled with the so-called peak-background split (PBS) approach. PBS is based on two assumptions~\citep{bardeen_1986,efstathiou_1988,cole_kaiser_1989,mo_white_1996,sheth_tormen_1999}: first, the abundance of halos depends on the amplitude of the matter power spectrum only through the variance, $\sigma^2(M)$, of the (linear) matter density field; second, the linear density contrast threshold for halo formation is unchanged by the presence of a long-wavelength density perturbation (the so-called called background field), i.e. a perturbation on scales much larger than the ones involved in the halo collapse. Given these two assumptions, according to PBS one could derive the halo density field and the halo bias parameters from the HMF~\citep{bardeen_1986,efstathiou_1988,cole_kaiser_1989,mo_white_1996,sheth_tormen_1999}, provided that the proper HMF is adopted.

Note that, since the multiplicity function describes halo formation in Lagrangian space, i.e. in the linearly evolved density field, the resulting halo number density changes according to the evolution of the long-wavelength density perturbation from the initial to the fully evolved (Eulerian) density field. Assuming that the dynamics of the long-wavelength perturbation is in the single-stream regime, the number of halos within the initial (Lagrangian) volume, $V^{\rm L}$, and final (Eulerian) volume, $V$, associated to the long-wavelength perturbation, $\delta_{\rm lw}$, is conserved. Therefore, mass conservation implies $V(z) [ 1 + \delta_{\rm lw}(z)] = V^{\rm L}(z_{\rm i}) [1 + \delta_{\rm lw}^{\rm L}D(z_{\rm i})/D(z=0)]$, where $z_{\rm i}$ is the initial redshift such that $D(z_{\rm i})/D(z=0)\ll 1$, and $\delta_{\rm lw}^{\rm L}$ is the Lagrangian field, i.e. the initial field linearly evolved to $z=0$ by definition.
The mapping from the Eulerian, $\delta_{\rm lw}(z)$, to the corresponding Lagrangian density contrast field, $\delta_{\rm lw}^{\rm L}$, is obtained by comparing the linear and nonlinear evolution of a spherical perturbation as explained e.g. on pp. 632-633 of~\cite{peebles_1993} and in Appendix A of~\cite{SVdW}.
\added{However there could be contributions which go beyond the spherical approximation \citep{ohta_2004,ohta_2003,sheth_2013}, and represent a second order effect in the Lagrangian-Eulerian mapping.
Nonetheless, for the values of the density contrast, scales and void ellipticity \citep{schuster_2022} considered in this work, the effect of non-spherical symmetry can be considered negligible. Moreover, any residual effect would be reabsorbed in the fitting of the $q$-parameter against the DEMNUni simulations. Therefore, we consider the assumption of spherical symmetry of the long-wavelength perturbation to be an accurate description of the Lagrangian-Eulerian mapping in the context considered in this work \citep{SVdW,jennings2013,massara2018}.}
It follows that the Eulerian HMF is modified by the long-wavelength perturbation according to
\begin{equation}\label{eq:eulerian_HMF}
\frac{{\rm d} n_{\rm h}}{{\rm d} M}= [1 + \delta_{\rm lw}(z)] \frac{\bar{\rho}_{\rm m}}{M} f[\tilde{\nu}(z), p, q] \frac{{\rm d} \tilde{\nu}(z)}{{\rm d} M}.
\end{equation}
This quantity is evaluated under the substitutions $\delta_{\rm c} \rightarrow \delta_{\rm c} - \delta_{\rm lw}^{\rm L}$, $\nu \rightarrow \tilde{\nu}=[\delta_{\rm sc}(z) - \delta_{\rm lw}^{\rm L} D(z=0)/D(z)] / \sigma(M)$, and $\bar{\rho}_{\rm m} \rightarrow \bar{\rho}_{\rm m} [1 + \delta_{\rm lw}(z)]$ in Eq.~\eqref{eq:universal_HMF}. Eq.~\eqref{eq:eulerian_HMF} represents the conditional HMF in the limit where the short-wavelength modes forming halos can be considered independent of long-wavelength modes~\citep{sheth_tormen_1999}.

For the first time to date, in this work we apply this approach to model the halo mass function and bias within cosmic voids. We treat the void shells as independent universes of almost constant background density, model the HMF within them, and accordingly, the halo density field as a function of the distance from the void center.
Since the HMF is a differential quantity, in order to estimate it, we divide the halo catalog into mass bins, $\Delta M$, and measure the number of halos, $N_{\rm h}(M,M+\Delta M)$, with mass between $M$ and $M+\Delta M$. Then, for each bin, we compare the measured differential number density of halos, $N_{\rm h}/V$, against the theoretical mass function integrated over the mass bin, i.e.
\begin{equation}\label{eq:measured_HMF}
\resizebox{\columnwidth}{!}{$
\begin{split}
&{\scriptstyle \rm SIMULATED-DATA:}&&{\cal N}_{\rm h}^{\rm SD}(M,\Delta M)\equiv\frac{N_{\rm h}(M,M+\Delta M)}{V}&\\
&{\scriptstyle \rm THEORY:}&&{\cal N}_{\rm h}^{\rm T}(M,\Delta M)\equiv\int_M^{M+\Delta M} \frac{{\rm d} n_{\rm h}(M',p,q)}{{\rm d} M'} {\rm d} M'.
\end{split}
$}
\end{equation}
To this aim, first we measure the HMF over the entire volume, $V$, of the comoving snapshots, i.e. the simulated Universe, where the background field is given by $\bar{\rho}_{\rm m}$, i.e. $\delta_{\rm lw}=0$. Adopting Eq.~\eqref{eq:universal_HMF} as HMF model, we implement a Markov Chain Monte Carlo (MCMC) analysis to evaluate the best-fit $p$ and $q$ parameters in the Universe, hereafter $p_{\rm U}$ and $q_{\rm U}$, finding a very good agreement with the simulated data. Then, we measure the HMF within each void shell, $\mathcal{S}^{\rm v}(r)$, i.e. along their density profiles. To model these measurements, we consider, as background field within voids, the total matter density contrast, $\bar{\rho}_{\rm m} [1+\delta^{\rm v}_{\rm m}(r,z)]$, i.e. we assume $\delta_{\rm lw}=\delta^{\rm v}_{\rm m}(r,z)$\footnote{The density contrast within voids is spherically symmetric as a result of the void stacking technique and of the isotropy of the Universe.} in Eq.~\eqref{eq:eulerian_HMF} and perform a new MCMC analysis to fit the $p$ and $q$ parameters to the simulated data. We find that they are different from $p_{\rm U}$ and $q_{\rm U}$, and, most importantly, are a function of the redshift and the distance, $r$, from the voids center. 
It follows that, to recover the agreement between simulated-data and theory in Eq.~\eqref{eq:measured_HMF}, and parameterize the dynamics of the halo collapse in underdense regions, the HMF model within voids, Eq.~\eqref{eq:eulerian_HMF}, requires $p$ and $q$ to be $r$- and $z$-dependent. 
Hereafter, $p_{\rm v}(r,z)$ and $q_{\rm v}(r,z)$ stand for $p$ and $q$ fitted to the HMF measured along void profiles, and we assume that our final model of the HMF inside cosmic voids is
\begin{align}\label{eq:HMF_voids}
\hspace{-1.2em}\frac{{\rm d} n_{\rm h}}{{\rm d} M}(\tilde{\nu},p_{\rm v}, q_{\rm v}, &\delta^{\rm v}_{\rm m}) 
= \sqrt{\frac{2}{\pi}}\frac{\bar{\rho}_{\rm m}}{M} (1 + \delta^{\rm v}_{\rm m})\times \\\nonumber
&\left\{ A_{\rm v} [ 1 + (q_{\rm v} \tilde{\nu}^2)^{-p_{\rm v}}] 
\sqrt{q_{\rm v}} e^{ -q_{\rm v}\tilde{\nu}^2/2 } \frac{{\rm d} \tilde{\nu}}{{\rm d} M}\right\}
\end{align}
where $A_{\rm v}=\left[ 1 + \Gamma \left( 1/2 -p_{\rm v} \right)/(2^{p_{\rm v}}\sqrt{\pi}) \right]^{-1}$. Here $p_{\rm v}(r,z)$ and $q_{\rm v}(r,z)$ effectively account for a possible correlation between the halo and void fields.

In Fig.~\ref{fig:HMF_in_shell} we show the goodness of Eq.~\eqref{eq:HMF_voids} in reproducing the measurements from the DEMNUni simulations. In particular, we present measurements of $\mathcal{N}_{\rm h}^{\rm SD}$ (symbols) compared to theoretical predictions $\mathcal{N}_{\rm h}^{\rm T}$ (lines), as defined in Eq.~\eqref{eq:measured_HMF}, at redshifts $z=0,\,0.49, \, 1.05$, for two different cases: i) the HMF of the entire comoving box and ii) the HMF within some representative spherical void shells, $\mathcal{S}^{\rm v}(r)$, with $r/R_{\rm eff}=0.23,0.37,0.5$. The bars represent the Poissonian uncertainty of the HMF measurements in $\mathcal{S}^{\rm v}(r)$ (blue) and in the box (red). The lines represent ${\cal N}_{\rm h}$ obtained using: i) $p_{\rm U}$ and $q_{\rm U}$ in Eq.~\eqref{eq:universal_HMF}, i.e. ${\cal N}_{\rm h}^{\rm T}(p_{\rm U},q_{\rm U},0)$ (red dash-dotted lines); ii) $\delta_{\rm lw}=\delta^{\rm v}_{\rm m}(r)$ together with $p_{\rm U}$ and $q_{\rm U}$ in Eq.~\ref{eq:eulerian_HMF}, i.e. ${\cal N}_{\rm h}^{\rm T}(p_{\rm U},q_{\rm U},\delta^{\rm v}_{\rm m})$ (orange dotted lines); iii) Eq.~\eqref{eq:HMF_voids}, i.e. ${\cal N}_{\rm h}^{\rm T}(p_{\rm v},q_{\rm v},\delta^{\rm v}_{\rm m})$ (blue solid lines); 
iv) and finally using $\delta_{\rm lw}=\delta^{\rm A}_{\rm m}(r)$ in Eq.~\eqref{eq:HMF_voids}, together with the corresponding $p^{\rm A}_{\rm v}(r)$ and $q^{\rm A}_{\rm v}(r)$ inferred via a MCMC analysis from the simulated HMF measurements, i.e. ${\cal N}_{\rm h}^{\rm T}(p^{\rm A}_{\rm v},q^{\rm A}_{\rm v},\delta^{\rm A}_{\rm m})$ (magenta dashed lines).
We stress how the last two approaches are able to recover with high accuracy the HMF along the void profiles, as shown from the residuals in the subplots of Fig.~\ref{fig:HMF_in_shell}. In addition, we verify the HMF behavior in ${\cal S}(r)$ is not reproduced if we use \textit{mean} $\bar{p}$ and $\bar{q}$ values measured within spheres of a given radius from the void center.
\begin{figure}[t!]
\centering
\includegraphics[width=\columnwidth]{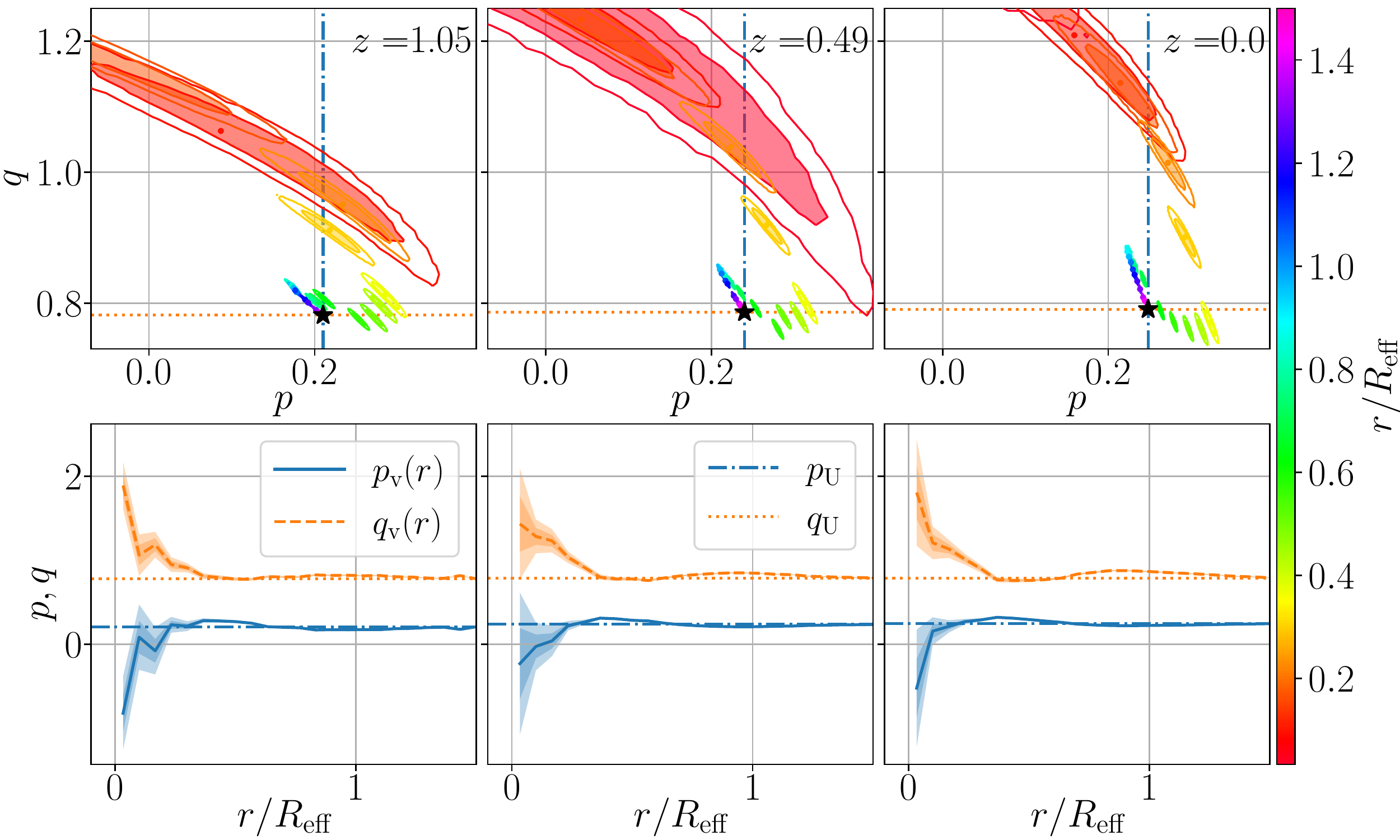}
\caption{The $p_{\rm v}(r)$ and $q_{\rm v}(r)$ parameters, along the void profile, of the multiplicity function associated to our HMF model in Eq.~\eqref{eq:HMF_voids}, fitted to simulated measurements in stacked void shells, for voids traced by halos with $M\geq 10^{13} h^{-1}M_\odot$ at $z=1.05,\,0.49,\,0$. The upper panels show the 68\% and 95\% CL in the $p$-$q$ plane. The contour colors denote the distance from the void center according to the colorbar on the right, and the black stars, together with the dot-dashed blue and dotted orange lines, represent $p_{\rm U}$ and $q_{\rm U}$. The lower panels show the same quantities as a function of $r$: $p_{\rm v}(r)$ is represented by the blue solid lines, $q_{\rm v}(r)$ by the orange dashed lines, $p_{\rm U}$ by the blue dash-dotted horizontal lines, and $q_{\rm U}$ by the orange dotted horizontal lines. Note that the $p$ and $q$ axis in the lower panels have a different scale with respect to the upper panels.}
\label{fig:pq_meas}
\end{figure}
\begin{figure*}[t!]
\centering
\includegraphics[width=\textwidth]{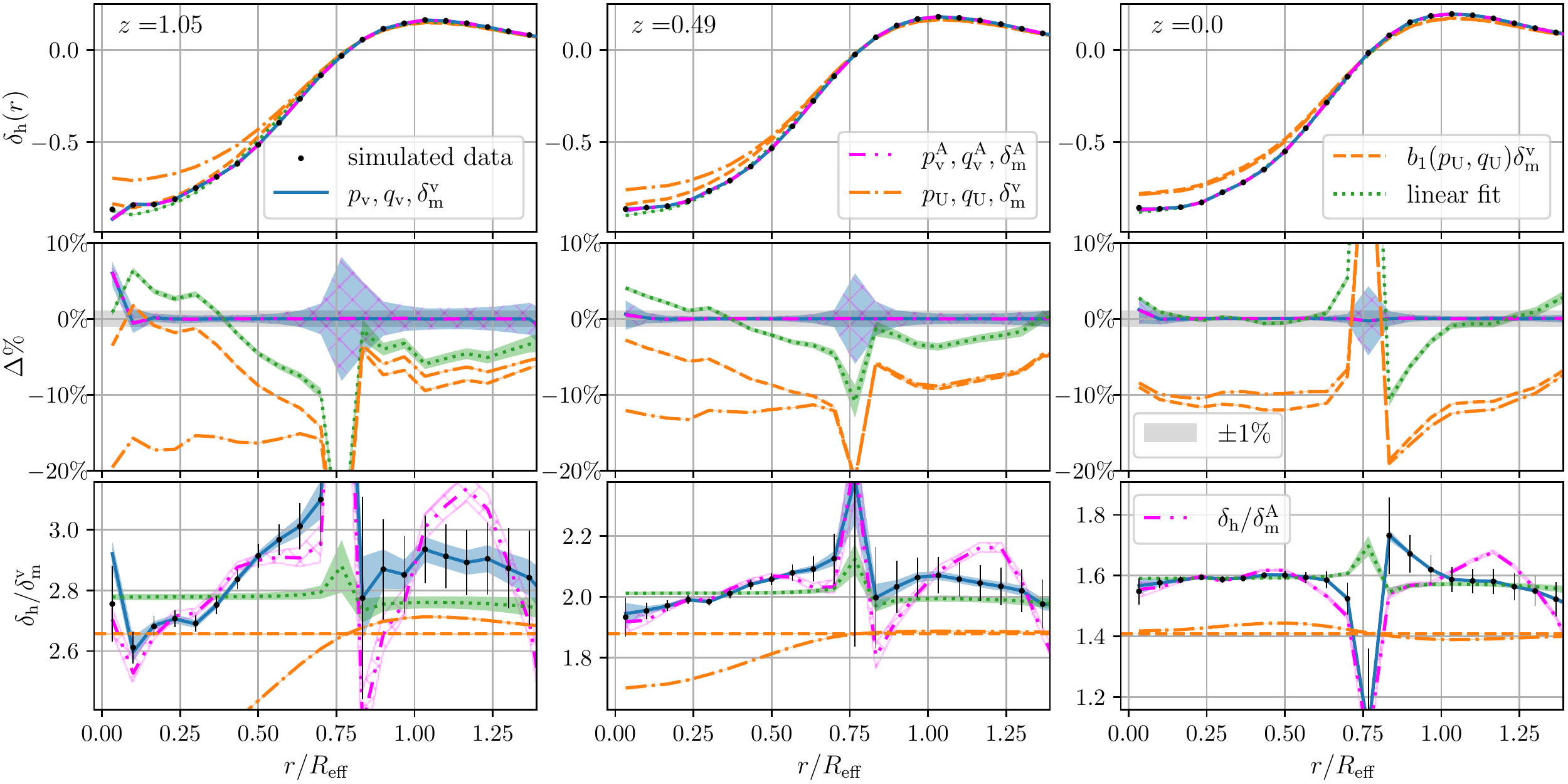}
\caption{The profile, $\delta_{\rm h}(r)$, for voids traced by halos with $M \geq 10^{13} h^{-1}M_\odot$ at $z=0,\,0.49,\,1.05$. 
The upper panels show the measured (black dots) halo density profile, $\delta^{\rm v}_{\rm h}(r)$, and the models described in the text: $p_{\rm v}(r),\,q_{\rm v}(r),\,\delta^{\rm v}_{\rm m}(r)$ is case a) (blue solid lines); $p^{\rm A}_{\rm v}(r),\,q^{\rm A}_{\rm v}(r),\,\delta^{\rm A}_{\rm m}(r)$ is case b) (magenta dash-dot-dotted lines); $p_{\rm U},\,q_{\rm U},\,\delta^{\rm v}_{\rm m}(r)$ is case c) (orange dash-dotted lines); $b_1(p_{\rm U},q_{\rm U})\delta_{\rm}^{\rm v}(r)$ is case d) (orange dashed lines); the linear fit is case e) (green dotted lines). The middle panels show the relative values of the model with respect to the measurements. The shaded areas represent the uncertainty of the parameters which the models depend on. The lower panels show the ratios $\delta_{\rm h}(r)/\delta_{\rm m}^{\rm v}(r)$, except for the magenta dash-dot-dotted lines which represent $\delta_{\rm h}(r)/\delta_{\rm m}^{\rm A}(r)$.}
\label{fig:bias_reconstr}
\end{figure*}

Fig.~\ref{fig:pq_meas} shows the parameters $p_{\rm v}(r)$ and $q_{\rm v}(r)$ fitted to the simulated data in each shell $\mathcal{S}^{\rm v}(r)$ at redshifts $z=0,\,0.49, \, 1.05$. In the upper panels we show how these parameters move in the $p$-$q$ plane following a continuous, $S$-shaped, path. At large $r$ they converge to $p_{\rm U}$ and $q_{\rm U}$ represented by the black stars, i.e. $p_{\rm U}=0.21,\,0.24,\,0.25$ and $q_{\rm U}=0.78,\,0.79,\,0.79$ at $z=1.05,\,0.49,\,0$, respectively. We find that this behavior is reproduced at all the redshifts and minimum halo masses considered. In the lower panels we show $p_{\rm v}(r)$ and $q_{\rm v}(r)$ as a function of the distance $r$ from the void center, together with $p_{\rm U}$ and $q_{\rm U}$ as blue dash-dotted and orange dotted lines, respectively. Here we do not show $p^{\rm A}_{\rm v}(r)$ and $q^{\rm A}_{\rm v}(r)$, as they mostly overlap with $p_{\rm v}(r)$ and $q_{\rm v}(r)$. In addition, even if in this work we present results only for $R_{\rm eff} \geq 2.5 \times {\rm mhs}$, we have verified that the same trend is preserved also for minimum $R_{\rm eff}$ values spanning from 1 to $3\times{\rm mhs}$, and for the  minimum halo masses listed in Tab.~\ref{tab:halo_mps}.

The trend observed in Fig.~\ref{fig:pq_meas} gives insight into the dynamics of halo collapse \textit{within voids}, effectively parameterized by $p_{\rm v}(r)$ and $q_{\rm v}(r)$, with $q=1$ and $p=0$ corresponding to the spherical collapse model~\citep{sheth_tormen_2002}. The $q$ parameter is determined by the number of massive halos~\citep{sheth_tormen_2001}. Fig.~\ref{fig:pq_meas} shows that $q_{\rm v}$ approaches and exceeds unity in the innermost void shells, suggesting that massive halos become increasingly rare nearing the void center. However, contrary to a first intuition, the blue curve in Fig.~\ref{fig:HMF_in_shell} shows that this trend is not only due to inner regions becoming increasingly underdense toward the center, i.e. to the dominant contribution of $\delta_{\rm m}^{\rm v}(r)$ with fixed $p=p_{\rm U}$ and $q=q_{\rm U}$ (orange solid curves), but also to the variation of $p=p_{\rm v}(r,z)$ and $q=q_{\rm v}(r,z)$ along the void profile, producing a smaller, although important effect, especially in the innermost regions.
The $p$ parameter is related to the shape of the moving barrier, and vanishes for a constant (spherical collapse) barrier~\citep{sheth_tormen_2001}. At each redshift considered, Fig.~\ref{fig:pq_meas} shows that, as the distance from the center increases, $p_{\rm v}$ oscillates around the value $p_{\rm v}\sim0.26$, being $p_{\rm v}<p_{\rm U}$ for $r<0.6 R_{\rm eff}$. This may suggest a smaller ellipticity in the collapse dynamics in the inner rather than outer void regions, though $p_{\rm v}(r,z)$ never approaches the $p=0$ value. Finally, let us notice how the trends in Figs.~\ref{fig:HMF_in_shell}--\ref{fig:pq_meas} show that, within voids, structure formation is slower and never reaches that of the Universe, behaving the dotted-orange line in Fig.~\ref{fig:HMF_in_shell} as an insurmountable barrier. This result may explain the findings in~\cite{tavasoli_2015}, i.e. that in sparse voids (as the inner regions in the top panels of Fig.~\ref{fig:HMF_in_shell}), galaxies are less massive and may be going through relatively slower evolution and continuing star formation.

\section{Halo bias in voids}
The way halos and galaxies trace the underlying matter distribution inside cosmic voids has been widely explored in the literature~\citep{furlanetto_piran_2006,hamaus_2014_vg_xc,sutter_2013,sutter_2014_DMgVoids,neyrinck_2014,chan_voidClustering_2014,clampitt_2015,pollina_2017,pollina_2019,contarini_2019,voivodic_2020}. For practical purposes, the halo density contrast, $\delta^{\rm v}_{\rm h}$, within voids is usually modeled~\citep{pollina_2017} assuming the following linear relationship with respect to the underlying dark matter density contrast, $\delta^{\rm v}_{\rm m}$
\begin{equation}\label{eq:linear_fit}
\delta_{\rm h}(r) = b_{\rm slope} \, \delta^{\rm v}_{\rm m} (r) + c_{\rm offset}\,,
\end{equation}
where, for watershed voids, the constant $c_{\rm offset}$ is generally consistent with zero within error margins, or assumed to be negligible~\citep{pollina_2017,pollina_2019,hamaus_2020,hamaus_2022_euclid}.
The assumption of a linear relation between halo and dark matter distributions is an empirical approach which parameterizes our ignorance about the way halos populate voids.

Here we show that a theoretical approach is feasible via the PBS formalism which, given a model for the HMF, provides a way to predict the halo density contrast, and therefore the halo bias~\citep{cole_kaiser_1989,bernardeau_1992,mo_jing_white_1997,sheth_tormen_1999}. In particular, we apply the PBS formalism inside voids using Eq.~\eqref{eq:HMF_voids} to compute the number density of halos with mass $M>M_{\rm min}$ within the void volume, $\mathcal{S}^{\rm v}(r)$, and Eq.~\eqref{eq:universal_HMF} to compute the number density of halos with mass $M>M_{\rm min}$ in the Universe. Then we define our model for the halo density contrast inside voids as
\begin{equation}\label{eq:halo_delta_pbs}
\delta_{\rm h}(r) = \frac{\int_{M_{\rm min}}^\infty \frac{{\rm d} n_{\rm h}}{{\rm d} M} \left[\tilde{\nu},p_{\rm v}(r),q_{\rm v}(r),\delta^{\rm v}_{\rm m}(r)\right]{\rm d} M}{\int_{M_{\rm min}}^\infty \frac{{\rm d} n_{\rm h}}{{\rm d} M}\left(\nu,p_{\rm U},q_{\rm U},0\right){\rm d} M} - 1 \,. 
\end{equation}
In the following we show that Eq.~\eqref{eq:halo_delta_pbs} is a theoretical relation able to reproduce with high accuracy the measured halo bias, in void shells, $\mathcal{S}^{\rm v}(r)$. It can be Taylor expanded in power of $\delta^{\rm v}_{\rm m}$ to get the bias expansion in $\mathcal{S}^{\rm v}(r)$, though, contrary to the $p$ and $q$ parameters in the literature, here $p_{\rm v}$ and $q_{\rm v}$ should not be kept constant with $r$. The expansion, truncated at the first order, and for $r\rightarrow\infty$, i.e. $p_{\rm v}(r)\rightarrow p_{\rm U}$ and $q_{\rm v}(r)\rightarrow q_{\rm U}$, reduces to the well known linear bias relation, which is supposed to be valid for $|\delta_{\rm h}| \sim |\delta^{\rm v}_{\rm m}| \ll 1$. 
However, note that this condition is not usually satisfied for density contrasts typical of voids, and in fact, we show below that the linear bias relation does not capture the halo bias trend in voids. A polynomial expansion would provide a more representative description~\citep{cole_kaiser_1989,bernardeau_1992,mo_jing_white_1997,fosalba_gaztanaga_1998,sheth_tormen_1999}. However, in this work we prefer not to use polynomial expansions and we implement the exact solution of Eq.~\eqref{eq:halo_delta_pbs}.

In Fig.~\ref{fig:bias_reconstr} we validate our Eq.~\eqref{eq:halo_delta_pbs} against the halo density contrast measurements (black dots) of the DEMNUni void profiles at $z=0,\,0.49,\,1.05$. We compare the following models: a) Eq.~\eqref{eq:halo_delta_pbs} (solid blue line); b) Eq.~\eqref{eq:halo_delta_pbs} using $p^{\rm A}_{\rm v}(r)$, $q^{\rm A}_{\rm v}(r)$ and $\delta^{\rm A}_{\rm m}(r)$  (dash-dot-dotted magenta line);
c) Eq.~\eqref{eq:halo_delta_pbs} using $p_{\rm U}$, $q_{\rm U}$ and $\delta^{\rm v}_{\rm m}(r)$ (dash-dotted orange line); d) the theoretical first order bias expansion of c), i.e. the linear bias relation described above, $\delta_{\rm h}(r) = b_1(p_{\rm U},q_{\rm U})\,\delta_{\rm m}^{\rm v}(r)$, where $b_1(p_{\rm U},q_{\rm U})$ is the linear term of the bias expansion (dashed orange line); e) the linear fit of Eq.~\eqref{eq:linear_fit} obtained by fitting $b_{\rm slope}$ and $c_{\rm offset}$ according to~\citet{pollina_2017} (dotted green line). The upper and middle panels show that cases a) and b) reproduce the simulated measurements along the whole void profile with relative uncertainty below 1\%, validating our theoretical approach, Eq.~\eqref{eq:halo_delta_pbs}. Model c) has a poor match to the simulated data. Note that the difference between model c) and model a) in Fig.~\ref{fig:bias_reconstr} is due to the variation of $p_{\rm v}$ and $q_{\rm v}$ which is important along the entire void profile and not only in the inner region, as shown in the top panel Fig.~\ref{fig:pq_meas}. Both the linear bias model d) and the empirical linear fit e) reproduce the void profile with an accuracy of $\sim 5-10\%$ depending on $z$. This can be explained considering that, as mentioned above, voids are nonlinearly evolved structures and, therefore, linear perturbation approaches may fail to describe the halo distribution within them. Moreover, the mismatch between approaches d) and e) could be considered a further evidence that, within voids, we are beyond the validity of the linear bias approximation and the exact solution of Eq.~\eqref{eq:halo_delta_pbs} combined with Eq.~\eqref{eq:HMF_voids} should be adopted. The lower panels show the ratios $\delta_{\rm h}(r)/\delta_{\rm h}^{\rm v}(r)$ at different $z$. This quantity helps visualize the \textit{scale-dependence} of the halo bias inside voids, which increases at larger redshifts, where nonlinear effect are more important, and is not fully captured by models d) and e). 
The dash-dot-dotted lines show the ratio $\delta_{\rm h} (r)/ \delta_{\rm m}^{\rm A}(r)$ for the analytical model of case b). The discrepancy with respect to the simulated data is due to the $\delta_{\rm m}^{\rm A}(r)$ fit, which is more accurate in the inner void underdense regions than at the compensation wall, $r/R_{\rm eff} \gtrsim 0.75$.\footnote{Note that the discontinuities at  $r/R_{\rm eff} \sim 0.75 $ in the middle and lower panels of Fig.~\ref{fig:bias_reconstr} are simply produced by numerical effects due to ratios between zero-crossing profiles.}

Finally, we stress that, using in Eq.~\eqref{eq:halo_delta_pbs} the analytical matter density profile $\delta^{\rm A}_{\rm v}(r)$, we are able to provide a theoretical model which predicts the halo bias in voids with 1\% accuracy with respect to the simulated data.

\section{Conclusions}
For the first time to date, in this work we have shown that the halo mass function inside cosmic voids is not universal, rather it depends on the distance, $r$, from the void center. We have verified this finding both via measurements of simulated data and a theoretical approach which exploits the PBS method, here novely adapted to the case of cosmic voids.

Eq.~\eqref{eq:HMF_voids} is one of the two main results of this work, and represents our theoretical model able to describe the HMF along void profiles, as it reproduces measurements from the halo-traced void catalogs identified with \vide in the DEMNUni cosmological simulations. In particular, in Fig.~\ref{fig:HMF_in_shell} we show that the HMF within voids depends on the distance, $r$, both via the matter density contrast and the parameters $p_{\rm v}$ and $q_{\rm v}$ that parameterize a multiplicity function having the \ST functional form.

In addition, applying the PBS technique to Eq.~\eqref{eq:HMF_voids}, we have been able to obtain the second main result of this work, Eq.~\eqref{eq:halo_delta_pbs}. As shown in Fig.~\ref{fig:bias_reconstr}, it provides a theoretical prediction of the halo bias within voids which reproduces with an accuracy of $\lesssim 1\%$ the measurements from the DEMNUni simulations. In this respect, we have also shown how the linear bias approximation fails to reproduce the correct trend, with $r$ and $z$, of halo bias measurements within voids.
Moreover, using the analytical matter density profile $\delta_{\rm m}^{\rm A}(r)$ \citep{hamaus_2014_profiles}, we have provided a theoretical model to predict the halo bias in voids.

Our theoretical modeling of the HMF and halo bias may have several applications in data analyses exploiting the relation between the halo and dark matter distributions inside voids, as for instance measurements of the void size function\footnote{Note that in this case the radius of each void is different from the watershed radius.}~\citep{contarini_2019,verza_2019,euclid_vsf}. 
A further application could be represented by redshift-space distortions around voids, that in ongoing and upcoming galaxy surveys will provide impressive power in constraining the growth of structures and the expansion history of the Universe ~\citep{hamaus_2020,nadathur_2020,hamaus_2022_euclid}. Here, the observable is the void profile traced by the galaxy distribution, therefore an accurate halo bias modeling inside voids would help improve the accuracy of this kind of analyses.
In addition, our bias model could be further improved via advanced analytical matter profiles to be fitted together with $p_{\rm v}$ and $q_{\rm v}$ against data.

Finally, the results presented in this work give insight into the physics of halo formation and halo bias in voids as they show how these depend on the distance from the void center.
Therefore, in a future work we will explore the relation between halo ellipticities and the $p_{\rm v}$ and $q_{\rm v}$ parameters within voids, as well as extend our modeling to cosmologies alternative to the \lcdm model, in particular in the presence of massive neutrinos and dynamical dark energy which may alter halo formation inside voids. An approach similar to the one we presented here could be also applied to the luminosity function of galaxies, exploring its relations with the halo mass function and the properties of galaxy populations in voids~\citep{peebles_2001,rojas_2004,vanDeWeygaert_2011,ricciardelli_2014,tavasoli_2015,beygu_2017,panchal_2020,habouzit_2020}.

\vspace{1.5em}

\noindent We warmly acknowledge Nico Hamaus, Alice Pisani, and Benjamin D. Wandelt for very useful comments. GV and AR are supported by the project ``Combining Cosmic Microwave Background and Large Scale Structure data: an Integrated Approach for Addressing Fundamental Questions in Cosmology'', funded by the MIUR Progetti di Rilevante Interesse Nazionale (PRIN) Bando 2017 - grant 2017YJYZAH.
AR acknowledges funding from Italian Ministry of Education, University and Research (MIUR) through the `Dipartimenti di eccellenza' project Science of the Universe. 
The DEMNUni simulations were carried out in the framework of ``The Dark Energy and Massive-Neutrino Universe'' project, using the Tier-0 IBM BG/Q Fermi machine and the Tier-0 Intel OmniPath Cluster Marconi-A1 of the
Centro Interuniversitario del Nord-Est per il Calcolo Elettronico (CINECA). We acknowledge
a generous CPU and storage allocation by the Italian Super-Computing Resource Allocation
(ISCRA) as well as from the HPC MoU CINECA-INAF, together with storage from INFN-CNAF and INAF-IA2.

\bibliography{biblio}{}
\bibliographystyle{aasjournal}

\listofchanges

\end{document}